\documentclass
[a4paper,aps,pra,twocolumn,amsmath,amssymb]{revtex4}

\usepackage{epsfig}
\usepackage{bm}
\usepackage{wasysym}

\usepackage{bbm}
\usepackage{amsmath}
\usepackage{verbatim}
\usepackage{graphicx}
\usepackage{dsfont}

\usepackage{marvosym} 

\newcommand{\mec}{{m}}
\newcommand{\omm}{\omega_\mec}
\newcommand{\am}{a_\mec}
\newcommand{\amd}{a_\mec^\dagger}
\newcommand{\Xm}{{X}_\mec}
\newcommand{\Pm}{P_\mec}
\newcommand{\gm}{\gamma_\mec^0}

\newcommand{\cav}{{c}}

\newcommand{\ac}{a_\cav}
\newcommand{\acd}{a_\cav^\dagger}
\newcommand{\xc}{X_\cav}
\newcommand{\pc}{P_\cav}

\newcommand{\om}[1]{\omega_{#1}}
\newcommand{\x}[1]{X_{#1}}
\newcommand{\p}[1]{P_{#1}}
\renewcommand{\a}[1]{a_{#1}}
\newcommand{\E}[1]{E_{#1}}

\renewcommand{\v}[1]{\vec{#1}}

\newcommand{\nm}{{\rm NM}}
\newcommand{\diag}{{\rm diag}}

\newcommand{\ket}[1]{|#1\rangle}

\newcommand{\In}{{\rm in}}
\newcommand{\Out}{{\rm out}}

\newcommand{\mean}[1]{\langle #1\rangle}

\newcommand{\kHz}{{\rm kHz}}

\begin{document}

\title{Observation of strong coupling between a micromechanical resonator\\and an optical cavity field}\thanks{This work was published in Nature \textbf{460}, 724 (2009).}
\author{Simon Gr\"oblacher$^{1,2}$, Klemens Hammerer$^{3,4}$, Michael R. Vanner$^{1,2}$, Markus Aspelmeyer$^{1,}$}\email{markus.aspelmeyer@quantum.at}
\affiliation{
$^1$ Institute for Quantum Optics and Quantum Information (IQOQI), Austrian Academy of Sciences, Boltzmanngasse 3, A-1090 Vienna, Austria\\
$^2$ Faculty of Physics, University of Vienna, Strudlhofgasse 4, A-1090 Vienna, Austria\\
$^3$ Institute for Quantum Optics and Quantum Information (IQOQI), Austrian Academy of Sciences, Technikerstra\ss e 21a, A-6020 Innsbruck, Austria\\
$^4$ Institute for Theoretical Physics, University of Innsbruck, Technikerstra\ss e 25, A-6020 Innsbruck, Austria}

\begin{abstract}
Achieving coherent quantum control over massive mechanical resonators is a current research goal. Nano- and micromechanical devices can be coupled to a variety of systems, for example to single electrons by electrostatic~\cite{Naik2006,Cleland2002} or magnetic coupling~\cite{Rugar2004,Rabl2009}, and to photons
by radiation pressure~\cite{Kippenberg2005,Gigan2006,Arcizet2006b,Thompson2008,Regal2008} or optical dipole forces~\cite{Eichenfield2007,Li2008}. So far, all such experiments have operated in a regime of weak coupling, in which reversible energy exchange between the mechanical device and its coupled partner is suppressed by fast decoherence of the individual systems to their local environments. Controlled quantum experiments are in principle not possible in such a regime, but instead require strong coupling. So far, this has been demonstrated only between microscopic quantum systems, such as atoms and photons (in the context of cavity quantum electrodynamics~\cite{Walther2006}) or solid state qubits and photons~\cite{Khitrova2006,Wallraff2004}. Strong coupling is an essential requirement for the preparation of mechanical quantum states, such as squeezed or entangled states~\cite{Bose1997,Marshall2003,Vitali2007,Clerk2008}, and also for using mechanical resonators in the context of quantum information processing, for example, as quantum transducers. Here we report the observation of optomechanical normal mode splitting~\cite{Marquardt2007,Dobrindt2008}, which provides unambiguous evidence for strong coupling of cavity photons to a mechanical resonator. This paves the way towards full quantum optical control of nano- and micromechanical devices.
\end{abstract}

\maketitle

A common feature of all coupled quantum systems is that their dynamics are dominated by the competition between the joint coupling rate and the rates at which the coupled systems decohere into their local environments. Only for sufficiently strong coupling can the effects of decoherence be overcome. This so-called "strong coupling
regime" is, in all cases, indispensable for the experimental investigation of a manifold of quantum phenomena. Nano- and micro-optomechanical oscillators are currently emerging as a new "textbook" example for coupled quantum systems. In this case, a single electromagnetic field mode is coupled to a (nano- or micrometre sized) mechanical oscillator. In analogy to cavity quantum electrodynamics (cQED), one can identify strong coupling as the regime where the coupling rate $g$ exceeds both the cavity amplitude decay rate $\kappa$ and the mechanical damping rate $\gamma_m$\textemdash as required, for example, in refs ~\cite{Bose1997,Marshall2003,Vitali2007}. Another class of proposals requires the weaker condition of "large cooperativity", that is, $g>\sqrt{\kappa\cdot\gamma_m}$ (refs~\cite{Clerk2008,Hammerer2009}). Strong coupling, ideally in combination with the preparation of zero entropy initial states (for example, by ground-state cooling of the mechanical resonator), is essential to obtain (quantum) control over this new domain of quantum physics. Whereas ground state preparation is a goal of continuing
research (in which much progress has been made, in particular by using optical laser cooling techniques~\cite{Wilson-Rae2008}), here we demonstrate strong optomechanical coupling using state-of-the-art micromechanical resonators.\\

Consider the canonical situation in which a mechanical resonator is coupled to the electromagnetic field of a high-finesse cavity via momentum transfer of the cavity photons (Fig.\ 1). The system naturally comprises two coupled oscillators:\ the electromagnetic field at cavity frequency $\omega_c$ (typically of the order of $10^{15}$~Hz) and the mechanical resonator at frequency $\omega_m$ ($\sim 10^7$~Hz). At first sight, the large discrepancy in the oscillator frequencies seems to inhibit any coupling; it is, however, alleviated by the fact that the cavity is driven by a laser field at frequency $\omega_L$, which effectively creates an optical oscillator at frequency $\Delta=\omega_c-\omega_L-\delta_{rp}$ (in a reference frame rotating at $\omega_L$; $\delta_{rp}$ is the mean shift of the cavity frequency due to radiation pressure). Each of the two oscillators decoheres into its local environment:\ the optical field at the cavity amplitude decay rate $\kappa$ and the mechanics at the damping rate $\gamma_m$. Entering the desired strong coupling regime requires a coupling rate $g\apprge\kappa,\gamma_ m$.\\

\begin{figure*}[htbp]
\centerline{\includegraphics[width=0.7\textwidth]{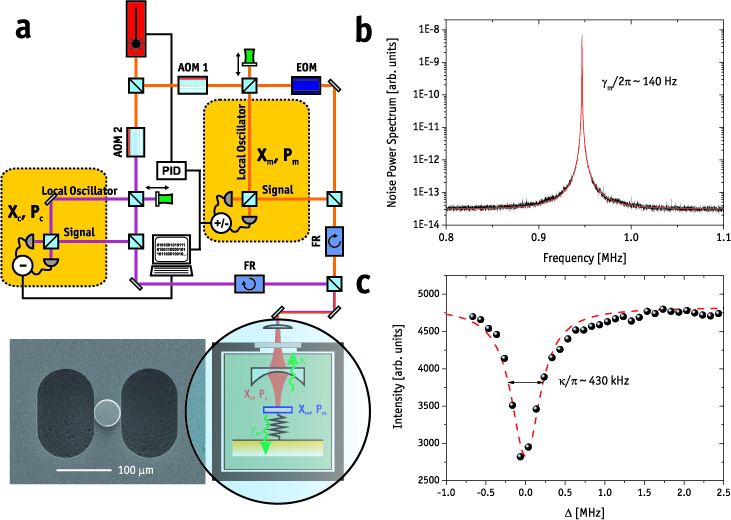}}
\caption{\textbf{Experimental set-up and characterization of the uncoupled mechanical and optical oscillator. a}, Our micromechanical resonator with a high-reflectivity mirror pad ($R>0.99991$) that forms the end-face of a 25-mm-long Fabry-P\'{e}rot cavity (magnified view circled, bottom right). A strong continuous-wave Nd:YAG laser is used to drive the optomechanical system (purple beam). By splitting off a faint part ($15~\mu$W) of the drive laser,
the laser frequency is actively locked to the Fabry-P\'{e}rot cavity frequency (orange beam). Locking is achieved by phase-modulation (electro-optical modulator, EOM) and by obtaining a Pound-Drever-Hall error signal required for feedback with a proportional-integral-derivative controller (PID). Acousto-optical modulators (AOM) control the relative frequency detuning $\Delta$ and thus allow for off-resonant driving of the cavity. Data presented here have been taken by varying the detuning $\Delta$ and the power of the drive beam. Both beams are coupled to the Fabry-P\'{e}rot cavity via the same spatial mode but orthogonal in polarization. The measured cavity linewidth (full-width at half-maximum, FWHM) $2\kappa\approx 2\pi\times 430$~kHz corresponds to an optical finesse $F\approx 14,000$. The fundamental mechanical mode of the microresonator at $\omega_m=2\pi\times 947$~kHz has a natural linewidth (FWHM) of $\gamma_m\approx 2\pi\times 140$~Hz (mechanical quality factor $Q\approx 6,700$) at room temperature. With $\kappa/\omega_m\approx 0.2$, these parameters place us well into the resolved sideband regime $\kappa/\omega_m\ll 1$. The effective mass of 145~ng was obtained by direct fitting of the optomechanical response at low driving powers. After interaction with the optomechanical system, both (drive and lock) beams are separated by a polarizing beamsplitter and Faraday rotators (FR) and are each independently measured by optical homodyning
(Supplementary Information). Each homodyne phase can be either scanned or locked to a fixed value by actuating a piezo-driven mirror. \textbf{b}, Mechanical noise power spectrum obtained by homodyne detection of the lock beam. Red line, fit to the data assuming an ideal harmonic oscillator in thermal equilibrium. \textbf{c}, Intensity of the drive beam that is reflected off the Fabry-P\'{e}rot cavity when scanning its detuning $\Delta$, which provides direct access to the cavity transfer function. Dashed red line, Lorentzian fit to the data.} \label{fig1}
\end{figure*}

The fundamental optomechanical radiation-pressure interaction $H_{int}=-\hbar g_0n_cX_m$ couples the cavity photon number $n_c$ to the position $X_m$ of the mechanics ($\hbar$ is $h/2\pi$, where $h$ is Planck's constant). On the
single-photon level, this interaction provides an intrinsically nonlinear coupling, where the coupling rate $g_0=\frac{\omega_c}{L}\sqrt{\frac{\hbar}{m\omega_m}}$ ($L$, cavity length; $m$, effective mass) describes the effect of a single photon on the optomechanical cavity. In all currently available optomechanical systems, however, $g_0$ is well below 100~Hz. Because the corresponding cavity decay rates are typically much larger than 10~kHz, the effect is too small to exploit the strong coupling regime on the single-photon level. For our experiment $g_0=2\pi\times 2.7$~Hz, which is smaller than both $\kappa$ ($2\pi\times 215$~kHz) and $\gamma_m$ ($2\pi\times 140$~Hz). To circumvent this limitation, we use a strong optical driving field ($\lambda=1,064$~nm), which shifts
the optomechanical steady state by means of radiation pressure from vacuum to a mean cavity amplitude $\alpha$ (mean cavity photon number $\langle n_c\rangle=\alpha^2$) and from zero displacement to a mean mechanical displacement $\beta$. The resulting effective interaction is obtained by standard mean-field expansion, and resembles two harmonic oscillators that are coupled linearly in their optical and mechanical position quadratures $X_c=(a_c+a_c^{\dag})$ and $X_m=(a_m+a_m^{\dag})$, respectively. This strongly driven optomechanical system is then described by equation (1) (see Supplementary Information):
\begin{equation}
H=\frac{\hbar\Delta}{2}(X_c^2+P_c^2)+\frac{\hbar\omega_m}{2}(X_m^2+P_m^2)-\hbar gX_cX_m
\end{equation}
The effective coupling strength $g=g_0\alpha$ is now enhanced by a factor of $\alpha=\sqrt{\langle n_c\rangle}$. Note that this enhancement comes at the cost of losing the nonlinear character of the interaction. Although there exist proposals that do require strong nonlinear coupling at the single-photon level~\cite{Marshall2003}, the majority of schemes for quantum optomechanical state manipulation work well within the regime of linear albeit strong coupling. They rely on the fact that linear interactions allow for protocols such as quantum state transfer and readout~\cite{Zhang2003}, generation of entanglement~\cite{Bose1997,Vitali2007}, conditional preparation of states via projective measurements on light~\cite{Clerk2008,Hammerer2009}, and so on, a fact which is well established in the fields of quantum optics and quantum information. In our experiment, by using external optical pump powers of up to 11~mW, we are able to achieve an increase in coupling by more than five orders of magnitude,
sufficient to reach the desired strong coupling regime.\\

\begin{figure*}[htbp]
\centerline{\includegraphics[width=0.6\textwidth]{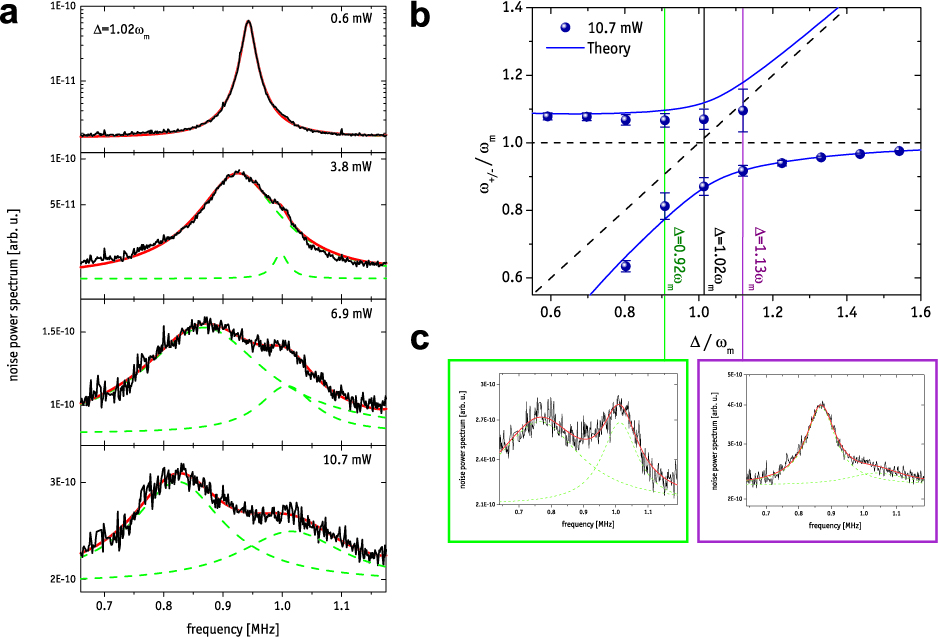}} \caption{\textbf{Optomechanical normal mode splitting and avoided crossing in the normal-mode frequency spectrum. a}, Emission spectra of the driven optomechanical cavity, obtained from sideband homodyne detection on the strong driving field after its interaction with the optomechanical system (see Supplementary Information). The power levels from top to bottom (0.6, 3.8,
6.9, 10.7~mW) correspond to an increasing coupling strength of $g=78$, 192, 260 and 325~kHz ($g=0.4$, 0.9, 1.2, 1.5~$\kappa$). All measurements are performed close to resonance ($\Delta=1.02\omega_m$). For strong driving powers a splitting of the cavity emission occurs, corresponding to the normal mode frequencies of true hybrid optomechanical degrees of freedom. This normal mode splitting is an unambiguous signature of the strong coupling regime. All plots are shown on a logarithmic scale. Green dashed lines are fits to the data assuming two independent Lorentzian curves, red solid lines are the sum signal of these two fits. \textbf{b}, Normal mode frequencies obtained from the fits to the spectra as a function of detuning $\Delta$. For far off-resonant driving, the normal modes approach the limiting case of two uncoupled systems. Dashed lines indicate the frequencies of the uncoupled optical (diagonal) and mechanical (horizontal) resonator, respectively. At resonance, normal mode splitting prevents a frequency degeneracy, which results in the shown avoided level crossing. Error bars, s.d. Solid lines are simulations (see Supplementary Information). For larger detuning values, the second normal mode peak could no longer be fitted owing to a nearby torsional mechanical mode. \textbf{c}, Normal mode spectra measured off resonance.} \label{fig2}
\end{figure*}

An unambiguous signature of strongly coupled systems is the occurrence of normal mode splitting, a phenomenon known to both classical and quantum physics. In the simplest case, two independent harmonic oscillators coupled via an additional joint spring will behave as a pair of uncoupled oscillators\textemdash so-called normal modes\textemdash with shifted resonance frequencies compared to the individual resonators. For the particular case of resonators with equal bare frequencies, a sufficiently strong coupling will introduce a spectral splitting of the two normal modes that is of the order of the coupling strength $g$. Normal mode splitting has been observed in a number of realizations of cQED, where it is also known as Rabi splitting, with photons coupled either to atoms~\cite{Thompson1992,Colombe2007,Aoki2006}, to excitons in semiconductor structures~\cite{Reithmaier2004,Weisbuch1992,Yoshie2004} or to Cooper pair box qubits in circuit QED~\cite{Wallraff2004}. In case of the strongly driven optomechanical system described by equation (1), the normal modes occur at frequencies $\omega_{\pm}^2=\frac{1}{2}(\Delta^2+\omega_m^2\pm\sqrt{(\Delta^2-\omega_m^2)^2+4g^2\omega_m\Delta})$ and exhibit a splitting $\omega_+-\omega_-\approx g$. In the given simple expression for normal mode frequencies, cavity decay and mechanical damping are neglected. A more careful analysis is carried out in the Supplementary Information, and shows that normal mode splitting occurs only above a threshold $g\apprge\kappa$ (refs~\cite{Marquardt2007,Dobrindt2008}) for our damped optomechanical system. The Hamiltonian can be re-written in terms of the normal modes and one obtains:
\begin{equation}
H=\frac{\hbar\omega_+}{2}(X_+^2+P_+^2)+\frac{\hbar\omega_-}{2}(X_-^2+P_-^2)
\end{equation}
For the resonant case $\Delta=\omega_m$, equation (2) describes two uncoupled oscillators with position and momentum quadratures $X_{\pm}=\sqrt{\frac{\omega_m\pm g}{2\omega_m}}(X_c\pm X_m)$ and $P_{\pm}=\sqrt{\frac{\omega_m}{2(\omega_m\pm g)}}(P_c\pm P_m)$. These new dynamical variables cannot be ascribed to either the cavity field or the mechanical resonator, but are true hybrid optomechanical degrees of freedom. The overall system energy spectrum $E_{m,n}$ is therefore given by the sum of the energies of the two normal modes, that is, $E_{m,n}=\hbar(m\omega_++n\omega_-)$. The degeneracy of the uncoupled energy levels is lifted, and normal mode splitting of adjacent levels occurs with a separation that is equivalent to the coupling strength $g$. In the presence of decoherence, the spectral lines are broadened to a width of ($\kappa+\gamma_m$) and the splitting can therefore only be resolved for $g\apprge\kappa,\gamma_m$, that is, for strong coupling.\\

We observe normal mode splitting via direct spectroscopy of the optical field emitted by the cavity. Emission of a cavity photon can in general be understood as a transition between dressed states of the optomechanical system, that is, between mechanical states that are dressed by the cavity radiation field. The structure of the optomechanical
interaction only allows for transitions that lower or raise the total number of normal mode excitations by one (see Supplementary Information). Photons emitted from the cavity therefore have to lie at sidebands equal to the dressed state frequencies $\omega_{\pm}$ relative to the incoming laser photons of frequency $\omega_L$, that is, they have to be emitted at sideband frequencies $\omega_L\pm\omega_+$ or $\omega_L\pm\omega_-$. Homodyne detection provides us with direct access to the optical sideband spectrum, which is presented in Fig.\ 2a for the resonant case $\Delta\approx\omega_m$. For small optical pump power, that is, in the regime of weak coupling, the splitting cannot be resolved and one obtains the well known situation of resolved sideband laser cooling, in which Stokes and anti-Stokes photons are emitted at one specific sideband frequency. The splitting becomes clearly visible at larger pump powers, which is unambiguous evidence for entering the strong coupling regime. Indeed, at a maximum optical driving power of $\sim 11$~mW, we obtain a coupling strength $g=2\pi\times 325$~kHz, which is larger than both $\kappa=2\pi\times 215$~kHz and $\gamma_m=2\pi\times 140$~Hz and which corresponds to the magnitude of the level crossing shown in Fig.\ 2b. As is expected, for detunings $\Delta$ off resonance, the normal mode frequencies
approach the values of the uncoupled system.\\

These characteristics of our strongly driven optomechanical system are reminiscent of a strongly driven two-level atom, and indeed a strong and instructive analogy exists. If an atom is pumped by a strong laser field, optical transitions can only occur between dressed atomic states, that is, atomic states "dressed" by the interaction with
the laser field. For strong driving, any Rabi splitting that is induced by strong coupling is effectively of order $G_0\sqrt{\langle n_L\rangle}$ ($n_L$, mean number of laser photons; $G_0$, electric dipole coupling) and one therefore obtains an equally spaced level splitting, fully analogous to the coupled optomechanical spectrum. From this point of view, the optomechanical modes can be interpreted in a dressed state approach as excitations of
mechanical states that are dressed by the cavity radiation field. The origin of the sideband doublet as observed in the output field of the strongly driven optomechanical cavity corresponds to the resonance fluorescence spectrum of a strongly driven atom, in which strong coupling gives rise to the two side-peaks in the so-called Mollow triplet. It is interesting to note that the analogy even holds for the single-photon regime, in which both systems are close to their quantum ground state. For both cases (that is, the atom-cavity system and the cavity-optomechanical system), a sufficiently strong single-photon interaction $g_0$ would allow one to obtain the well known vacuum Rabi splitting as well as state-dependent level spacing, which is due to intrinsic nonlinearities in the coupling.\\

We should stress that normal mode splitting alone does not establish a proof for coherent dynamics, that is, for quantum interference effects. With the present experimental parameters, such effects are washed out by thermal decoherence and normal mode splitting has a classical explanation in the framework of linear dispersion theory~\cite{Zhu1990}. Still, the demonstration of normal mode splitting is a necessary condition for future quantum experiments.\\

We finally comment on the prospects for mechanical quantum state manipulation in the regime of strong coupling. One important additional requirement in most proposed schemes is the initialization of the mechanical device close to its quantum ground state. Recent theoretical results show that both ground state laser cooling and strong coupling can be achieved simultaneously, provided that the conditions $\frac{k_BT}{\hbar Q}\ll\kappa\ll\omega_m$ are fulfilled~\cite{Dobrindt2008,Wilson-Rae2008}. Thus, in addition to operating in the resolved sideband regime, a thermal decoherence rate that is small compared to the cavity decay rate is required. Cryogenic experiments have demonstrated thermal decoherence rates as low as 20~kHz for nanomechanical resonators for a 20~mK environment temperature~\cite{Regal2008}. For our experiment, temperatures below 300~mK would be sufficient to combine strong coupling with ground state cooling.\\

We have demonstrated strong coupling of a micromechanical resonator to an optical cavity field. This regime is a necessary precondition to obtaining quantum control of mechanical systems. Together with the availability of high-quality mechanical resonators operated at low temperatures, which minimizes thermal decoherence of the mechanics, strong optomechanical coupling provides the basis for full photonic quantum control of massive mechanical resonators. We suggest that future developments will eventually also allow strong coupling to be achieved in the nonlinear regime, that is, at the single-photon level~\cite{Li2008,Marshall2003}, to exploit optomechanical vacuum Rabi splitting.\\

\newpage
\begin{center}
\Large{\textbf{Supplementary Information}}
\end{center}

\normalsize{
\section{Normal Mode Frequencies and Damping Rates}}

As shown in \cite{Wilson-Rae2007,Marquardt2007,Genes2008} the linearized Hamiltonian for a driven cavity mode coupled via radiation pressure to a harmonically bound mirror is
\begin{align}\label{Eq:Hamiltonian}
H=\frac{\hbar\Delta}{2}(\xc^2+\pc^2)+\frac{\hbar\omm}{2}(\Xm^2+\Pm^2)-\hbar g\xc\Xm
\end{align}
with an opto-mechanical coupling rate $g=g_0\alpha=\frac{2}{L}\sqrt{\frac{P\kappa\omega_c}{m\omega_\mec(\kappa^2+\Delta^2)}}$ (following \cite{Genes2008}) for an input power $P$ of the driving laser ($L,~\omega_c$ and $\kappa$ are cavity length, resonance and amplitude decay rate respectively, $m$ the effective mass of the mechanical oscillator). For a two-sided cavity with decay rate $\kappa$ through the input-coupler and $\bar{\kappa}$ through the oscillating mirror, this formula generalizes to $g=\frac{2}{L}\sqrt{\frac{P\kappa\omega_c}{m\omega_\mec((\kappa+\bar\kappa)^2+\Delta^2)}}$.

It is convenient to define $\v{R}^T=(\xc,\pc,\Xm,\Pm)$ and express the Hamiltonian as $H=\frac{\hbar}{2}\v{R}^TM\v{R}$ where
\[
M=\begin{pmatrix}
  \Delta &   0 & g & 0 \\
    0 & \Delta & 0 & 0 \\
    g &   0 & \omm & 0 \\
    0 &   0 & 0 & \omm
  \end{pmatrix}.
\]
The transformation to normal modes $\v{R}^\nm=(\x+,\p+,\x-,\p-)$ is achieved with a linear transformation $\v{R}^\nm=S\v{R}$, where $S$ fulfills $M=S^T\diag(\om+,\om+,\om-,\om-)S$ and is symplectic, i.e. it obeys $J=SJS^T$ where
\[
J=
\begin{pmatrix}
  0 & 1 &  0 & 0 \\
 -1 & 0 &  0 & 0 \\
  0 & 0 &  0 & 1 \\
  0 & 0 & -1 & 0 \\
\end{pmatrix}.
\]
The latter property guarantees that canonical commutation relations are conserved, i.e. $[\v{R}_i,\v{R}_j]=[\v{R}^\nm_i,\v{R}^\nm_j]=iJ_{ij}$. The explicit form of $S$ can in principle be determined, but is quite involved and does not give much insight. As will become clear in a moment, the normal mode frequencies $\om\pm$ can be easily calculated without constructing $S$ and are given by
\begin{align}\label{Eq:omega}
\om\pm^2=\frac{1}{2}\left(\Delta^2+\omm^2\pm\sqrt{(\Delta^2-\omm^2)^2+4g^2\omm\Delta}\right).
\end{align}

The canonical operators evolve according to
\begin{align}\label{Eq:EoM}
  \dot{\v{R}}(t)&=i[H,\v{R}(t)]-D\v{R}(t)-\sqrt{2D}\v{R}_\In(t)\nonumber\\
  &=(JM-D)\v{R}(t)-\sqrt{2D}\v{R}_\In(t),
\end{align}
where we included damping of the cavity field and the mechanical resonator with $D={\rm diag}(\kappa,\kappa,\gm,\gm)$  and Langevin forces $\v{R}_\In(t)=(x_\In,p_\In,f_{\Xm},f_{\Pm})$. For white vacuum noise input to the cavity and a thermal white noise bath coupling to the mechanical system, all first moments vanish $\mean{\v{R}(t)}\equiv 0$ and the only non-zero two time correlation functions are
\begin{align}\label{Eq:NoiseCorrs}
  \mean{x_\In(t)x_\In(t')}&=\mean{p_\In(t)p_\In(t')}=\frac{1}{2}\delta(t-t'),\nonumber\\
  \mean{f_{\Xm}(t)f_{\Xm}(t')}&=\mean{f_{\Pm}(t)f_{\Pm}(t')}=\left(\bar{n}+\frac{1}{2}\right)\delta(t-t'),
\end{align}
where $\bar{n}\simeq\frac{kT}{\hbar\omm}$.

\begin{figure*}[t]
\begin{center}
\includegraphics[width=1\textwidth]{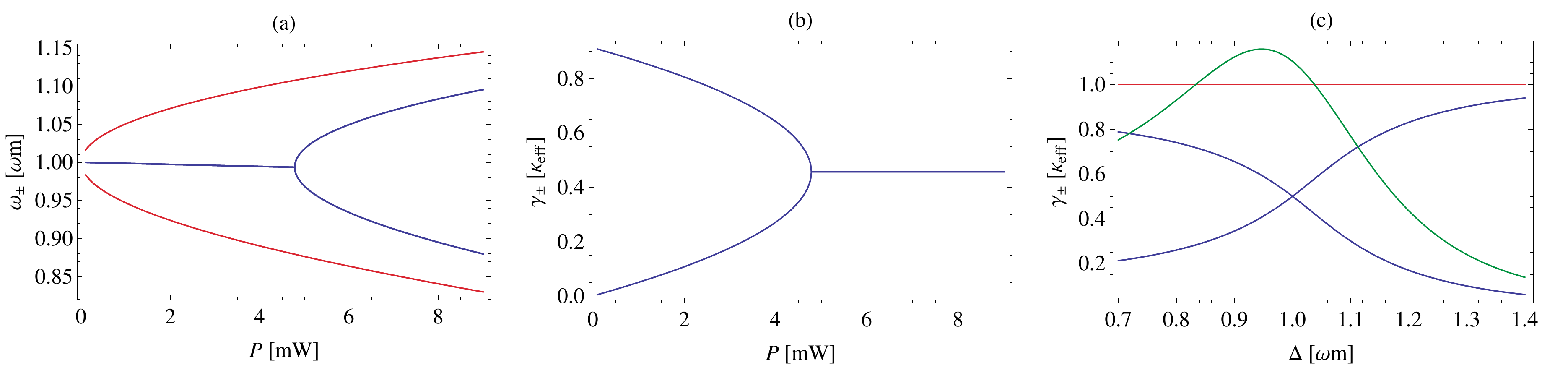}
\end{center}
\caption{(a) Normal mode frequencies $\om{\pm}$ for undamped system (red) and damped system (blue) for varying power of driving laser. (b) Same for effective normal mode damping $\gamma_\pm$. (c) Effective damping rates of normal modes (blue), cavity amplitude decay rate $\kappa$ (red) and effective mechanical decay rate $\gamma_\mec$ (green) for varying detuning. Not shown is the natural mechanical damping rate as $\gm/\kappa\simeq 10^{-3}$. Parameters are as in the main text, $\omm=2\pi\times 947~\kHz$, $\gm=2\pi\times140~{\rm Hz}$, $m=145~{\rm ng}$, $L=2.5~{\rm cm}$, $\om{c}=1.77\times10^{15}~{\rm Hz}$, $\kappa=2\pi\times172~\kHz$ and $\bar\kappa=2\pi\times43~\kHz$. In (a) and (b) $\Delta=\omm$ and in (c) $P=10.7~{\rm mW}$.}\label{Fig:NMFrequencies}
\end{figure*}

From Eq.~\eqref{Eq:EoM} it is clear that eigenfrequencies and effective damping rates of the system are given by, respectively, the imaginary and real parts of the eigenvalues of $i(JM-D)$. The eigenvalues occur in complex conjugate pairs and the imaginary parts of the ones in the upper half plane determine eigenfrequencies. For the undamped system, $D=0$, the eigenvalues are purely complex and one arrives at expression \eqref{Eq:omega} for the normal mode frequencies. For the damped system, $D\neq0$, the eigenvalues of $i(JM-D)$ will in general be complex and thus determine normal mode frequencies $\omega_\pm$ and effective damping rates $\gamma_\pm$ of normal modes, as exemplified in Fig.~\ref{Fig:NMFrequencies}. The theoretical values of the normal mode frequencies $\omega_\pm$ in Fig.~2a of the main text were as well determined in this way. While normal mode splitting (NMS) occurs for any non-zero coupling $g$ in an undamped system, a threshold of $g\gtrsim\kappa$ has to be surpassed to observe NMS \cite{Marquardt2007,Dobrindt2008}. Effective damping rates behave complementary and merge above the same threshold.  Comparison of the normal mode damping rates $\gamma_\pm$ to the effective mechanical damping rate $\gamma_\mec=\gm+\frac{2g^2\kappa\Delta\omm}{[\kappa^2+(\Delta-\omm)^2][\kappa^2+(\Delta+\omm)^2]}$, as derived in the theory for cavity-assisted cooling \cite{Genes2008,Wilson-Rae2007}, shows that the condition for resolving the normal mode peaks is $g\gg\kappa,\gamma_\mec$.\\

\section{Cavity Emission Spectrum}

\begin{figure}[ht]
\begin{center}
\includegraphics[width=\columnwidth]{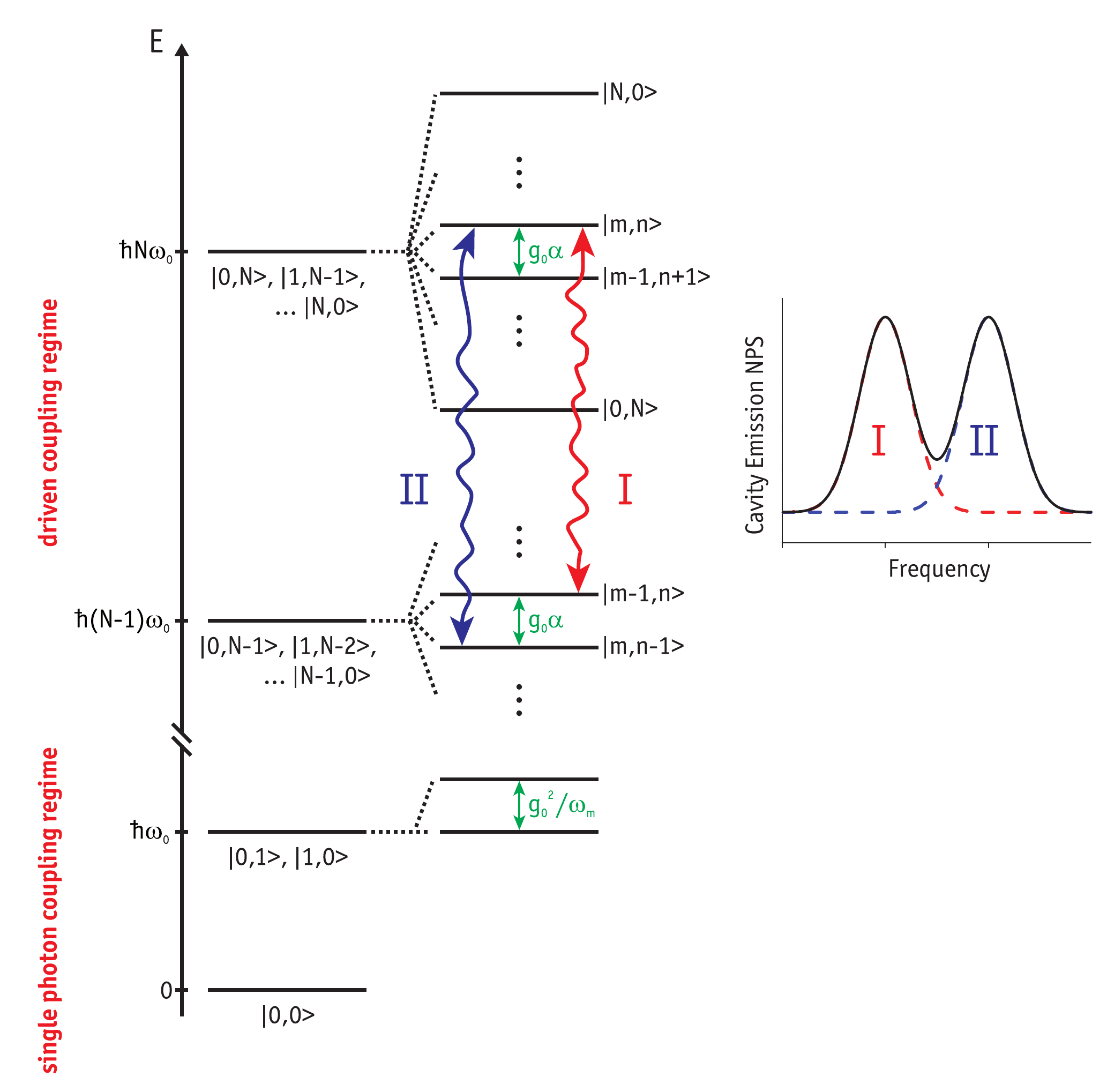}
\end{center}
\caption{Energy spectrum of a driven opto-mechanical cavity. For a degenerate, uncoupled system (left), $\omega_m=\Delta$, $g_0=0$, the spectrum consists of equidistant multipletts of energy $\hbar N\omega_m$ and degeneracy N ($\hbar$: Planck's constant; $N$: number of excitations; $\omega_m$: mechanical resonance frequency). For a coupled system (right), $g_0\neq 0$, the degeneracy is broken. In the strongly driven regime, where the cavity is in a coherent state with mean number of photons $\langle n_c\rangle$, the levels in each  $N$-multiplett split up by $g=g_0\sqrt{\langle n_c\rangle}$  into dressed states $\ket{m,n}$ with $m+n=N$. Emission of a cavity photon is accompanied only by transitions $\ket{m,n}\leftrightarrow\ket{m-1,n}$ or $\ket{m,n}\leftrightarrow\ket{m,n-1}$ between dressed states. Accordingly, emitted photons have to lie at sideband-frequencies $\omega_L+\omega_\pm$. This gives rise to a doublet structure in the sideband spectrum (bottom) with a splitting $\omega_+-\omega_-\approx g$. The observed normal-mode splitting is shown in Figure 2 of the main text. In the single photon coupling regime, the fundamentally anharmonic nature of the spectrum becomes important, with a splitting between dressed states scaling like $\frac{g_0^2 n_c^2}{\omega_m}$ (shown is the "opto-mechanical vacuum Rabi-splitting" for $n_c=1$). In the present experiment we cannot access this nonlinear regime, which would require a large single photon coupling $g_0\gg\kappa,\gamma_m$.}\label{Fig:DressedState}
\end{figure}

\subsection{Dressed States and Exact Diagonalization}

In terms of normal mode operators the linearized Hamiltonian \eqref{Eq:Hamiltonian} is given by $H=\frac{\hbar\om+}{2}(\x+^2+\p+^2)+\frac{\hbar\om-}{2}(\x-^2+\p-^2)$. It can be expressed also in terms of creation and annihilation operators $\a\pm=(\x\pm+i\p\pm)/\sqrt{2}$ as $H=\hbar\om+\left(\a+^\dagger\a++\frac{1}{2}\right)+\hbar\om-\left(\a-^\dagger\a-+\frac{1}{2}\right)$. The Eigenstates and -energies are thus $H\ket{n,m}=\E{n,m}\ket{n,m}$, where
\begin{align*}
  \ket{n,m}&=\frac{1}{\sqrt{n!m!}}(\a+^\dagger)^n(\a-^\dagger)^m\ket{0,0},\\
  \E{n,m}&=\hbar\om+(n+\frac{1}{2})+\hbar\om-(m+\frac{1}{2}).
\end{align*}

Emission of a cavity photon is in general accompanied by a transition of the opto-mechanical system from one eigenstate to another by changing a single excitation, $\ket{n,m}\leftrightarrow\ket{n-1,m}$ and $\ket{n,m}\leftrightarrow\ket{n,m-1}$.
The energy splitting between these states is $\E{n,m}-\E{n-1,m}=\hbar\om+$ and $\E{n,m}-\E{n,m-1}=\hbar\om-$ respectively. Photons emitted from the cavity have to carry away this energy excess/deficiency relative to the incoming laser photons of frequency $\om{L}$, i.e. they  have to have frequencies $\om{L}\pm\om+$ or $\om{L}\pm\om-$. Transitions between the dressed opto-mechanical states and the associated emission dublett is illustrated in Fig.\,\ref{Fig:DressedState}.

In order to compare the low-energy part of the opto-mechanical spectrum to the one of the Jaynes Cummings system, as shown in Fig.~4, we give here the exact eigenstates and -values of the non-linear radiation pressure Hamiltonian
\[
H=\hbar\omm\amd\am+\hbar\Delta\acd\ac-\hbar g_0\amd\am(\ac+\acd).
\]
It is straight forward to check that $H\ket{\psi_{k,n}}=E_{k,n}\ket{\psi_{k,n}}$ with
\begin{align*}
  \ket{\psi_{k,n}}&=\exp\left[\textstyle{\frac{g_0n}{\omm}}(\am-\amd)\right]\ket{k}_m\ket{n}_c, \\
  E_{k,n}&=\hbar\left(\omm k+\Delta n +\textstyle{\frac{g_0^2}{\omm}n^2}\right).
\end{align*}
That is, the eigenstates are shifted Fock states of the uncoupled system and the energy spectrum is anharmonic with a {\it quadratic} dependence in the photon number. The ''opto-mechanical Rabbi splitting'' is thus $\frac{g_0^2}{\omm}$, see also Fig.\,\ref{Fig:DressedState}.

\subsection{Emission Power Spectrum}

\begin{figure*}[t]
\begin{center}
\includegraphics[width=0.6\textwidth]{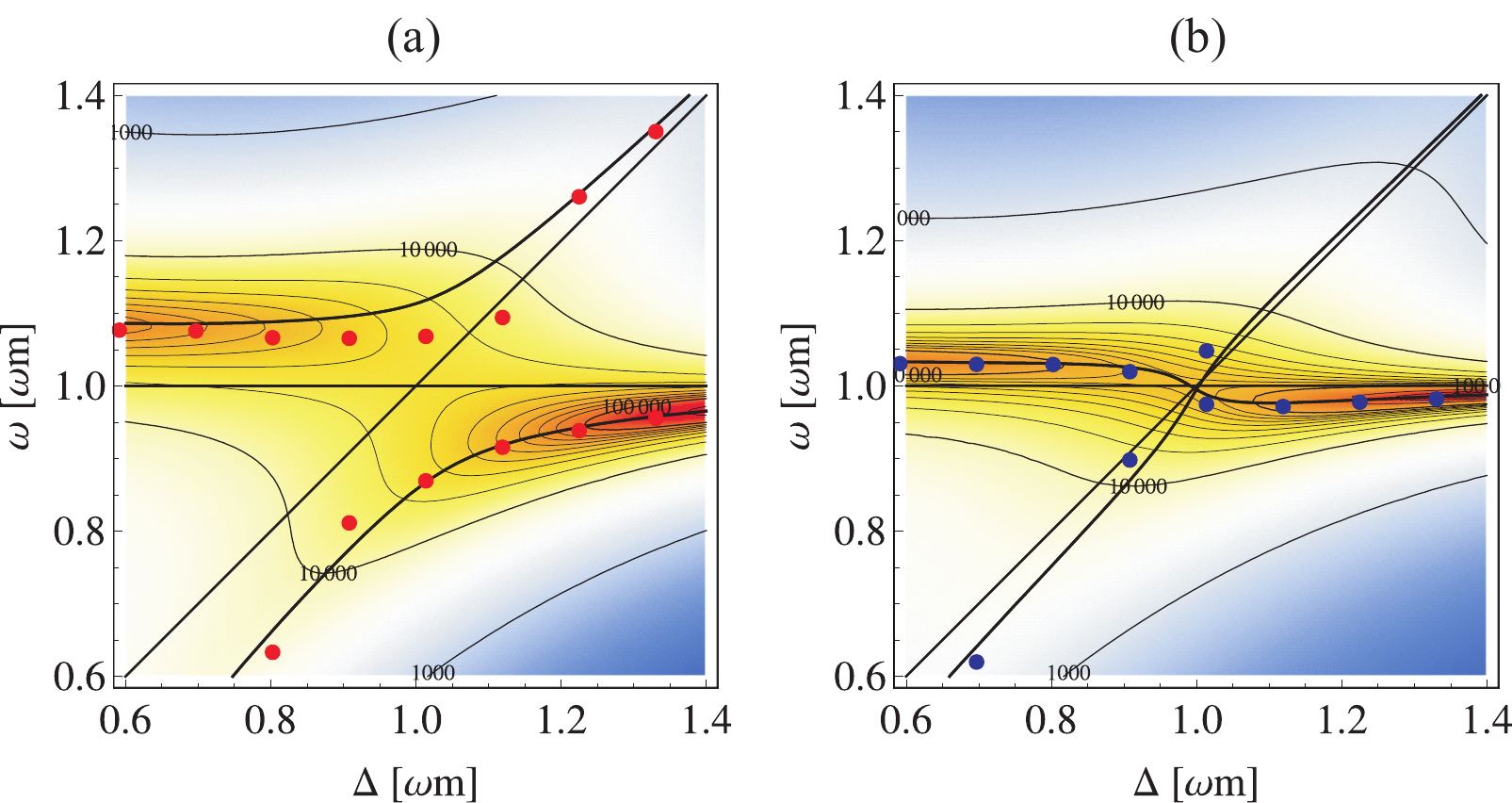}
\end{center}
\caption{(a) Emission spectrum $S_{NPS}(\omega)$ of opto-mechanical system for varying detuning (colors refer to a logarithmic scale of arbitrary units). Peak positions are well described by the normal mode eigenfrequencies $\om{\pm}$ (black lines). For comparison we reproduce also the measured data of presented in Fig.~3b of the main text. Parameters are as in \ref{Fig:NMFrequencies}, the input input power was $P=10.7~{\rm mW}$. Expression \eqref{Eq:Gamma} was used to evaluate the spectrum with a cavity decay rate through the input-coupler $\kappa=2\pi\times172~\kHz$ and at rate $\bar\kappa=2\pi\times43~\kHz$ through the oscillating mirror. (b) Same for a power value $P=3.8~{\rm mW}$ just below threshold for normal mode splitting, cf. Fig.~\ref{Fig:NMFrequencies}a.}\label{Fig:AvoidedCrossing}
\end{figure*}

The power spectral density of light emitted by the cavity is explicitly determined as follows: In frequency space [$\v{R}(\omega)=\int d\omega\v{R}(t)\exp(i\omega t)/\sqrt{2\pi}$] the steady state solutions to the equations of motion \eqref{Eq:EoM} are
\begin{align}\label{Eq:StatSol}
  \v{R}(\omega)=\frac{1}{i\omega+JM-D}\sqrt{2D}\v{R}_\In(\omega).
\end{align}
With the quantum optical cavity input-output relations it follows that
\begin{align*}
\v{R}_\Out(\omega)&=\sqrt{2D}\v{R}(\omega)+\v{R}_\In(\omega)\\
&=\left(\sqrt{2D}\frac{1}{i\omega+JM-D}\sqrt{2D}+1\right)\v{R}_\In(\omega),
\end{align*}
where $\v{R}_\Out(\omega)=(x_\Out,p_\Out,f_{\Xm,\Out},f_{\Pm,\Out})$. $(x_\Out,p_\Out)$ are quadratures for the cavity output field which are subject to homodyne detection. In order to calculate their stationary properties we formally introduce also ''phononic output fields'' $f_{\Xm,\Out},f_{\Pm,\Out})$. The spectral correlation functions can be collected in a Hermitean spectral $4\times4$ correlation matrix $\gamma^\Out_{ij}(\omega,\omega')=\mean{(\v{R}_\Out(\omega'))_i(\v{R}_\Out(\omega))_j}$.
Straight forward calculation yields $\gamma^\Out(\omega,\omega')=\delta(\omega+\omega')\Gamma(\omega)$ where
\begin{align*}
  \Gamma(\omega)&=\left(\sqrt{2D}\frac{1}{i\omega+JM-D}\sqrt{2D}+1\right)N\\
  &\quad\times\left(\sqrt{2D}\frac{1}{-i\omega+JM-D}\sqrt{2D}+1\right)^T
\end{align*}
and $N={\rm diag}\left(\frac{1}{2},\frac{1}{2},\bar{n}+\frac{1}{2},\bar{n}+\frac{1}{2}\right)$. If losses through the second mirror with amplitude decay rate $\bar\kappa$ are taken into account, the last expression generalizes to
\begin{align}\label{Eq:Gamma}
  \Gamma(\omega)&=\left(\sqrt{2D}\frac{1}{i\omega+JM-D-\bar{D}}\sqrt{2D}+1\right)N\nonumber\\
  &\quad\times\left(\sqrt{2D}\frac{1}{-i\omega+JM-D-\bar{D}}\sqrt{2D}+1\right)^T\nonumber\\
  &\quad+\left(\sqrt{2D}\frac{1}{i\omega+JM-D-\bar{D}}\sqrt{2\bar{D}}\right)\bar{N}\nonumber\\
  &\quad\times\left(\sqrt{2\bar{D}}\frac{1}{-i\omega+JM-D-\bar{D}}\sqrt{2D}\right)^T,
\end{align}
where $\bar{D}={\rm diag}(\bar{\kappa},\bar{\kappa},0,0)$ and $\bar{N}={\rm diag}(\frac{1}{2},\frac{1}{2},0,0)$.

Finally, the spectral density $S(\omega)$ is defined as $S(\omega)\delta(\omega+\omega')=\mean{a^\dagger_\Out(\omega')a^{\phantom\dagger}_\Out(\omega)}$ where the amplitude operator for the cavity output field is $a^{\phantom\dagger}_\Out(\omega)=(x_\Out(\omega)+ip_\Out(\omega))\sqrt{2}$. It follows from the definition of the spectral correlation matrix given above that
\begin{align*}
  S(\omega)=\frac{1}{2}\left[\Gamma_{11}(\omega)+\Gamma_{22}(\omega)+i(\Gamma_{12}(\omega)-\Gamma_{21}(\omega))\right].
\end{align*}
This expression gives the spectral density of sideband modes at a frequency $\om{L}+\omega$. In homodyne detection of sideband modes we do not distinguish sideband frequencies $\om{L}\pm\omega$ and extract only the overall noise power spectrum at a sideband frequency $|\omega|$, which is given by $S_{NPS}(\omega)=\sqrt{S(\omega)^2+S(-\omega)^2}$ and shown in Fig.~\ref{Fig:AvoidedCrossing}. The simple consideration in terms of dressed state transitions as given above shows good agreement with the exact calculated positions of spectral peaks, which are in turn in excellent agreement with measured data presented already in Fig.~2b of the main text.

\section{Homodyne detection of optical and mechanical quadratures}

We obtain the generalized optical and mechanical quadratures $X_c$ and $X_m$ via two independent, simultaneous optical homodyne measurements. Homodyne detection requires the mixing of a strong local oscillator field with the signal beam at a symmetric beamsplitter and a balanced photodetection at the beamsplitter output ports. The difference photocurrent then provides a direct measure of the generalized quadrature $X(\phi,t)=a(t)e^{i\phi}+a^{\dag}(t)e^{-i\phi}$ of an optical beam ($\phi$: local oscillator phase), where $X(\phi=0,t)$ and $X(\phi=\frac{\pi}{2},t)$ are the well-known amplitude and phase quadratures, respectively. To measure $X_c$, homodyning was performed on the driving beam after its interaction with the cavity. The second homodyning measures the locking beam after its resonant interaction with the cavity. Because of the weak interaction (we choose the power of the lock beam such that $g\ll\kappa$) the cavity field phase quadrature adiabatically follows the evolution of the mechanics and hence provides direct access to $X_m$. The local oscillator phase in the homodyne measurement of the locking field was always actively stabilized to detect the locking beam phase quadrature. Each of the two difference photocurrents was recorded independently by a high-speed analogue-to-digital converter (14 bit, 10 MSample sec$^{-1}$). The mechanical and optical noise power spectra from Figures 1b and 2a, respectively, were directly inferred from these recorded time traces. In that case, the local oscillator phase of the drive field was locked to a fixed value.

\begin{acknowledgements}
We are grateful to T. Corbitt, C. Genes, S. Go\ss ler, P. K. Lam, G. Milburn, P. Rabl and P. Zoller for discussions. We also thank M. Metzler, R. Ilic and M. Skvarla (CNF), and K. C. Schwab and J. Hertzberg, for microfabrication support, and R. Blach for technical support. We acknowledge financial support from the Austrian Science Fund FWF (projects P19539 and START), the European Commission (project MINOS) and the Foundational Questions Institute (project RFP 2-08-03). S.G. is a recipient of a DOC fellowship of the Austrian Academy of Sciences; S.G. and M.R.V. are members of the FWF doctoral programme Complex Quantum Systems (CoQuS).
\end{acknowledgements}

\end{document}